\documentclass[12pt]{article}
\usepackage{amsmath,amssymb}
\usepackage{lmodern}
\usepackage{xcolor}
\usepackage[margin=.8in]{geometry}
\usepackage{color}
\usepackage{fancyvrb}
\usepackage{graphicx}
\usepackage{calc}
\usepackage[inline]{enumitem}
\usepackage{graphicx}
\usepackage{array}
\usepackage[unicode]{hyperref}
\usepackage{url}
\usepackage{natbib}
\usepackage{authblk}

\newcommand\blfootnote[1]{%
  \begingroup
  \renewcommand\thefootnote{}\footnote{#1}%
  \addtocounter{footnote}{-1}%
  \endgroup
}

\title{A note on simulation methods for the Dirichlet-Laplace prior
\blfootnote{A slightly modified version of this article is included in Luis Gruber's master thesis \citep{gruber2025}. In the process of preparing this corrigendum, it has come to our attention that the issue was independently uncovered by \cite{onorati2025}.}
}
\author[1]{Luis Gruber}
\author[1]{Gregor Kastner}
\author[2]{Anirban Bhattacharya}
\author[3]{Debdeep Pati}
\author[4]{Natesh Pillai}
\author[5]{David Dunson}

\affil[1]{Department of Statistics, University of Klagenfurt}
\affil[2]{Department of Statistics, Texas A\&M University}
\affil[3]{Department of Statistics, University of Wisconsin--Madison}
\affil[4]{Department of Statistics, Harvard University}
\affil[5]{Department of Statistical Science, Duke University}

\date{This note appears in \emph{The Journal of the American Statistical Association}, \url{https://doi.org/10.1080/01621459.2025.2540256}.}

\begin{document}
\maketitle
\begin{abstract}
\citet{dl} introduce a novel prior, the
Dirichlet-Laplace (DL) prior, and propose a Markov chain Monte Carlo
(MCMC) method to simulate posterior draws under this prior in a
conditionally Gaussian setting. The original algorithm samples from
conditional distributions in the wrong order, i.e., it does not correctly sample from the joint posterior distribution of all latent variables.
This note details the issue and provides two simple solutions: A
correction to the original algorithm and a new algorithm based on an
alternative, yet equivalent, formulation of the prior. This corrigendum does not affect the theoretical results in \citet{dl}.
\end{abstract}

\section{Model, prior and proposed MCMC algorithm}
\label{the-model-the-prior-and-the-proposed-mcmc-algorithm}

Consider the standard normal means setting
\begin{equation}\label{eq:model}
y_i = \theta_i + \epsilon_i, \quad \epsilon_i \sim \mathcal{N}(0,1), \quad 1 \le i \le n.
\end{equation} In Bhattacharya et al. (2015), a prior distribution for
the vector of means,
\(\boldsymbol{\theta}=(\theta_1,\dots,\theta_n)^\prime\), is proposed.
It can be represented as the following Gaussian scale-mixture:
\begin{equation}\label{eq:prior}
\theta_j \sim \mathcal{N}(0,\psi_j \phi_j^2 \tau^2), \quad \psi_j \sim \mathcal{E}xp(1/2), \quad \boldsymbol{\phi} \sim \mathcal{D}ir(a,\dots,a), \quad \tau \sim \mathcal{G}(na, 1/2).
\end{equation} This prior has attractive theoretical properties, was
successfully applied to a range of practical and econometric problems \citep[e.g.,][]{zhang_bondell, kastner, bhapplications2022, hub-etal:int}, and consequently became a widely used continuous shrinkage prior.

To be of use for practical analysis, a posterior simulator via Markov
chain Monte Carlo (MCMC) is developed in \citet{dl}. It
aims at simulating draws from the joint posterior distribution
\(p(\boldsymbol{\theta},\boldsymbol{\phi},\tau,\boldsymbol{\psi}|\boldsymbol{y})\).
The algorithm is intended to mix well, and thus relies on blocking and
marginalization in order to reduce autocorrelation. The sampler cycles
through the following steps:

\begin{enumerate}
\def\labelenumi{\arabic{enumi}.}
\item
  \(\boldsymbol{\theta}|\boldsymbol{\psi}, \boldsymbol{\phi}, \tau, \boldsymbol{y}\):
  \(\theta_j \sim \mathcal{N}(\mu_j,\sigma^2_j)\) independently, with
  \(\sigma_j^2 = \left(1+\frac{1}{\psi_j \phi_j^2 \tau^2}\right)^{-1}\)
  and \(\mu_j = \sigma_j^2 y_j\), \(j=1,\dots,n\).
\item
  \(\boldsymbol{\psi}|\boldsymbol{\theta},\boldsymbol{\phi},\tau\):
  Sample
  \(\tilde{\psi_j}|\theta_j,\phi_j,\tau \sim iG(\frac{\phi_j\tau}{|\theta_j|}, 1)\)
  independently and set \(\psi_j=\frac{1}{\tilde{\psi_j}}\).
\item
  \(\tau|\boldsymbol{\phi},\boldsymbol{\theta} \sim giG(na-n,1, 2\sum_{j=1}^n\frac{|\theta_j|}{\phi_j})\).
\item
  \(\boldsymbol{\phi}|\boldsymbol{\theta}\): Sample
  \(T_j|\theta_j \sim giG(a-1,1,2|\theta_j|)\) independently and set
  \(\phi_j = \frac{T_j}{\sum_{j=1}^nT_j}\).
\end{enumerate}
It is important to note that a key feature is the joint update of
\(\boldsymbol{\psi},\boldsymbol{\phi},\tau|\boldsymbol{\theta}\).

\section{On the ordering of MCMC updates}\label{the-problem}

Joint updates, i.e., updates where the order of updating single elements
matters, are a common source of errors in MCMC algorithms. \citet{gir} draws direct attention to this problem; nevertheless, it has
since appeared undetected in other influential publications \citep[e.g.,][]{primiceri, primiceri_corrigendum}. Because joint updates
appear to be such a common source for errors and to facilitate
exposition, we first abstract the problem to a general level.

Assume the joint distribution \(p(d,e,f)\), where the full conditional
distributions \(p(d|e,f)\), \(p(e|d,f)\), and \(p(f|e,d)\) are of well
known forms. To generate \(M\) draws from \(p(d,e,f)\), one can
successively cycle through

\begin{enumerate}
\item $d^{(m)} \sim p(d|e^{(m-1)},f^{(m-1)})$, 
\item $e^{(m)} \sim p(e|d^{(m)},f^{(m-1)})$, 
\item $f^{(m)} \sim p(f|d^{(m)},e^{(m)})$,
\end{enumerate}
where the parenthesized superscript denotes the respective MCMC
iteration. The three updating steps are interchangeable, since they
consist of sampling from full conditional distributions.

Things get slightly more involved by marginalizing. Assume that the
conditional distribution \(p(e|f)\) -- marginalized over \(d\) -- is of
well-known form. Then, the product \(p(e|f)p(d|e,f)\) is the
marginal-conditional decomposition of the joint conditional distribution
\(p(d,e|f)\). A correct sampler, exploiting the joint update of
\(p(d,e|f)\), cycles through

\begin{enumerate}
\item $f^{(m)} \sim p(f|d^{(m-1)},e^{(m-1)})$, 
\item
\begin{enumerate}
\item[(a)] $e^{(m)} \sim p(e|f^{(m)})$, 
\item[(b)] $d^{(m)} \sim p(d|e^{(m)},f^{(m)})$.
\end{enumerate}
\end{enumerate}

In that case, the steps 2(a) and 2(b) are \emph{not} interchangeable. A correct algorithm samples \(e\) conditional on \(f\) -- marginalized over \(d\) -- and then samples \(d\) conditional on both \(e\) and \(f\) just drawn.
In short, to sample from the joint distribution of two variables, it is
possible to first draw from the marginal of one and then generate the
other conditionally on that draw, thereby following the principle
\textit{marginal first -- conditional later}.

Taking a closer look at the ingredients of the steps 2., 3., and 4., of
the algorithm proposed in \citet{dl}, one observes that
in the conditionals of both
\(\tau|\boldsymbol{\phi},\boldsymbol{\theta}\) and
\(\boldsymbol{\phi}|\boldsymbol{\theta}\), the auxiliary variable
\(\boldsymbol{\psi}\) is integrated out. In addition, the conditional
\(\boldsymbol{\phi}|\boldsymbol{\theta}\) is marginalized over \(\tau\).
Together with the full conditional
\(\boldsymbol{\psi}|\boldsymbol{\theta}, \boldsymbol{\phi}, \tau\),
those two conditionals form the marginal-conditional decomposition of
the joint conditional
\(\boldsymbol{\phi},\tau,\boldsymbol{\psi}|\boldsymbol{\theta}\):
\begin{equation}
p(\boldsymbol{\phi},\tau,\boldsymbol{\psi}|\boldsymbol{\theta}) = p(\boldsymbol{\phi},\tau|\boldsymbol{\theta}) p(\boldsymbol{\psi}|\boldsymbol{\theta}, \boldsymbol{\phi}, \tau) = p(\boldsymbol{\phi}|\boldsymbol{\theta})p(\tau|\boldsymbol{\phi},\boldsymbol{\theta}) p(\boldsymbol{\psi}|\boldsymbol{\theta}, \boldsymbol{\phi}, \tau).
\end{equation} Clearly, the proposed algorithm cycles through the
conditionals in the wrong order, because within the joint update of
\(\boldsymbol{\phi},\tau,\boldsymbol{\psi}|\boldsymbol{\theta}\) it does
not adhere to the principle
\textit{marginal first -- conditional later}.

\section{Two solutions}\label{two-solutions}

In this section, we present two solutions how to correctly sample from
the posterior distribution under the prior in Equation \ref{eq:prior}.
The first one is simply a correction to the original one. The second
builds on another -- yet equivalent -- formulation of the prior.

The corrected original algorithm cycles through the following steps:

\begin{enumerate}
\item $\boldsymbol{\theta}|\boldsymbol{\psi}, \boldsymbol{\phi}, \tau, \boldsymbol{y}$: $\theta_j \sim \mathcal{N}(\mu_j,\sigma^2_j)$ independently, with $\sigma_j^2 = \left(1+\frac{1}{\psi_j \phi_j^2 \tau^2}\right)^{-1}$ and $\mu_j = \sigma_j^2 y_j$, $j=1,\dots,n$.
\item For the joint update of $p(\boldsymbol{\phi},\tau,\boldsymbol{\psi}|\boldsymbol{\theta})$ sample
\begin{enumerate}
\item[(a)] $T_j|\theta_j \sim giG(a-1,1,2|\theta_j|)$ independently and set $\phi_j = \frac{T_j}{\sum_{j=1}^nT_j}$,
\item[(b)] $\tau|\boldsymbol{\phi},\boldsymbol{\theta} \sim giG(na-n,1, 2\sum_{j=1}^n\frac{|\theta_j|}{\phi_j})$,
\item[(c)] $\tilde{\psi_j}|\theta_j,\phi_j,\tau \sim iG(\frac{\phi_j\tau}{|\theta_j|}, 1)$ independently and set $\psi_j=\frac{1}{\tilde{\psi_j}}$.
\end{enumerate}
\end{enumerate}
Note that the ordering within the joint update of
\(p(\boldsymbol{\phi},\tau,\boldsymbol{\psi}|\boldsymbol{\theta})\) is
not interchangeable.

The second solution builds on an alternative formulation of the prior,
which is mentioned in Section 3 (Equation 10) in \citet{dl}. It uses the fact that
\(\lambda_j=\phi_j\tau \sim \mathcal{G}(a,1/2)\) independently for
\(j=1,\dots,n\): \begin{equation}\label{eq:prior2}
\theta_j \sim \mathcal{N}(0,\psi_j \lambda_j^2), \quad \psi_j \sim \mathcal{E}xp(1/2), \quad \lambda_j \sim \mathcal{G}(a, 1/2).
\end{equation} The conditional posterior distributions of the
alternative formulation are standard and hence not derived. The
algorithm associated with the alternative formulation cycles through the
following steps:
\begin{enumerate}
\item $\boldsymbol{\theta}|\boldsymbol{\psi}, \boldsymbol{\lambda}, \boldsymbol{y}$: $\theta_j \sim \mathcal{N}(\mu_j,\sigma^2_j)$ independently, with $\sigma_j^2 = \left(1+\frac{1}{\psi_j \lambda_j^2}\right)^{-1}$ and $\mu_j = \sigma_j^2 y_j$, $j=1,\dots,n$.
\item For the joint update of $p(\boldsymbol{\lambda},\boldsymbol{\psi}|\boldsymbol{\theta})$ sample
\begin{enumerate}
\item[(a)] $\lambda_j|\theta_j \sim giG(a-1,1,2|\theta_j|)$ independently,
\item[(b)] $\tilde{\psi_j}|\theta_j,\lambda_j \sim iG(\frac{\lambda_j}{|\theta_j|}, 1)$ independently and set $\psi_j=\frac{1}{\tilde{\psi_j}}$.
\end{enumerate}
\end{enumerate}

Both algorithms are intrinsically connected: The conditional posterior
distribution of \(T_j|\theta_j\) in the corrected algorithm, which is
based on Theorem 2.1 in \citet{dl}, is of the same
distribution as the conditional posterior distribution of
\(\lambda_j|\theta_j\) in the alternative algorithm. The explanation is
as follows: Since a Dirichlet-distributed vector always sums up to
unity, the following holds:
\(\sum_{j=1}^n \lambda_j = \sum_{j=1}^n\phi_j \tau = \tau\). That is,
\(\phi_j\) for \(j=1,\dots,n\) can be recovered with
\(\phi_j=\frac{\lambda_j}{\sum_{j=1}^n \lambda_j}\). Hence, it must be
that \(T_j \propto \lambda_j\).

We conclude this section by noting that a similar Gibbs sampler based on \eqref{eq:prior2} was employed by the authors to produce some of the results in the manuscript during the revision process, and also in \cite{pati2014posterior} -- the only difference being that $\lambda_j$ was sampled given $\theta_j$ as well as $\psi_j$ (i.e., no blocking was performed). Doing so avoided the need to sample from a generalized inverse Gaussian distribution, whose MATLAB implementation at the time was slow. 

\section{``Getting it right''}\label{gewekes-test}

In the following, we demonstrate that the proposed algorithm does not
generate draws from the distributions it aims at sampling from. Here, we
are applying a slight variation of ``Getting it right'' test proposed in \citet{gir}. In a nutshell: Let
\(\boldsymbol{\kappa}=( \boldsymbol{\psi}, \boldsymbol{\phi}, \tau)^\prime\)
denote the vector of hyperparameters. Then, one possibility to sample
from the joint distribution of
\(p(\boldsymbol{\theta},\boldsymbol{\kappa})\) is the
\textit{marginal-conditional simulator} (mcs): It first samples
\(\boldsymbol{\kappa}^{(m)} \sim p(\boldsymbol{\kappa})\), and then
\(\boldsymbol{\theta}^{(m)} \sim p(\boldsymbol{\theta}|\boldsymbol{\kappa}^{(m)})\).
The resulting sequence
\(\{\boldsymbol{\kappa}^{(m)}, \boldsymbol{\theta}^{(m)}\}\) is \(iid\).
Another way of sampling from the joint distribution of
\(p(\boldsymbol{\theta},\boldsymbol{\kappa})\) is the
\textit{successive-conditional simulator} (scs). It is initialized with
one single draw
\(\boldsymbol{\kappa}^{(0)} \sim p(\boldsymbol{\kappa})\). Then, it
successively iterates over
\(\boldsymbol{\theta}^{(m)} \sim  p(\boldsymbol{\theta}|\boldsymbol{\kappa}^{(m-1)})\) and
\(\boldsymbol{\kappa}^{(m)} \sim q(\boldsymbol{\kappa}|\boldsymbol{\kappa}^{(m-1)},\boldsymbol{\theta}^{(m)})\),
where \(q(\cdot)\) is the transition kernel of the posterior MCMC
algorithm with respect to the joint conditional posterior
\(p(\boldsymbol{\kappa}|\boldsymbol{\theta})\). Comparing the samples
from both simulators allows for checking the correctness of the joint
update.

\begin{figure}[tp]
\centering
\includegraphics[width=\textwidth]{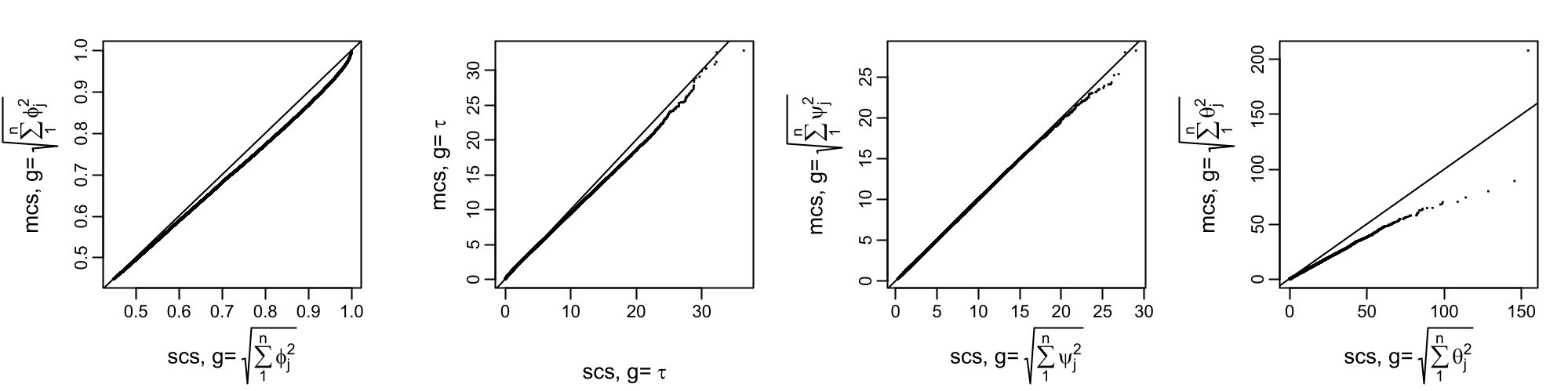}
\includegraphics[width=\textwidth]{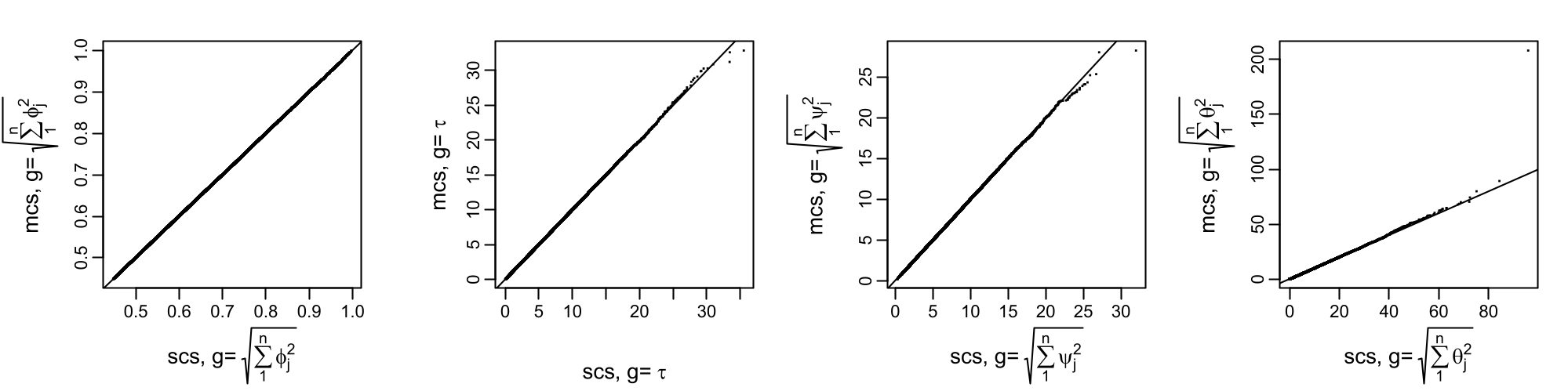}
\includegraphics[width=\textwidth]{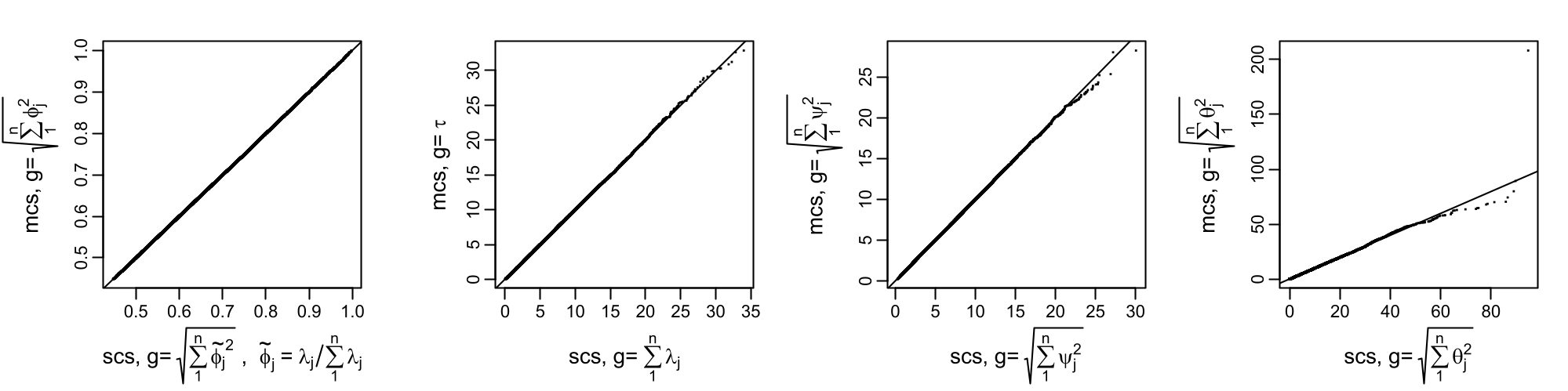} 
\caption{QQ-plots comparing samples from the \emph{marginal-conditional sampler} (mcs) and the \emph{successive-conditional sampler} (scs) with the original (top row), the corrected (middle row), and the alternative (bottom row) MCMC transition kernel. Plots are based on $250,000$ iterations of each simulator. The concentration parameter is $a=0.5$ and the dimensionality is $n=5$.}
\label{fig:geweke}
\end{figure}

The top row of Figure \ref{fig:geweke} depicts QQ-plots comparing samples from the
marginal-conditional simulator to samples from the
successive-conditional sampler with the transition kernel of the
original algorithm for several parameters of the
model.\footnote{Computer code for all simulators can be found at \url{https://github.com/gregorkastner/dl}.}
For almost all considered test functions \(g(\boldsymbol{\kappa})\), the
quantiles do not align.
The middle row of Figure \ref{fig:geweke} depicts QQ-plots for the corrected
algorithm. The quantiles of the successive-conditional sampler with the
corrected transition kernel align with the quantiles of the
marginal-conditional sampler.
The bottom row of Figure \ref{fig:geweke} depicts QQ-plots for the alternative
algorithm. It demonstrates that the transformations
\(g^{(m)}=\frac{\lambda_j^{(m)}}{\sum_{j=1}^n\lambda_j^{(m)}}\) and
\(g^{(m)}=\sum_{j=1}^n\lambda_j^{(m)}\) of the successive-conditional
simulator correspond to the sequences of \(\phi_j^{(m)}\) and
\(\tau^{(m)}\) of the marginal-conditional simulator, respectively.

\section{Empirical results}
\label{results-based-on-two-correct-algorithms}

\begin{table}[tp]
\caption{Squared error comparison over $100$ replicates. Average squared error across replicates reported for DL with $a=1/n$ and $a=1/2$ based on the original, the corrected, and the alternative algorithm. Signal strength $A=5$ and $A=6$.}
\label{tab:sim1}
\centering
\resizebox*{\textwidth}{!}{
\begin{tabular}[t]{rl!{\extracolsep{6pt}}r!{\extracolsep{0pt}} r !{\extracolsep{6pt}}r!{\extracolsep{0pt}} r!{\extracolsep{6pt}}r!{\extracolsep{0pt}} r !{\extracolsep{6pt}}r!{\extracolsep{0pt}} r !{\extracolsep{6pt}}r!{\extracolsep{0pt}} r !{\extracolsep{6pt}}r!{\extracolsep{0pt}} r}
\hline\hline
& \multicolumn{1}{r}{$n$}&\multicolumn{6}{c}{100} & \multicolumn{6}{c}{200}\\
\cline{3-8}\cline{9-14}
&\multicolumn{1}{r}{$\frac{q_n}{n}$\%}&\multicolumn{2}{c}{5}&\multicolumn{2}{c}{10}&\multicolumn{2}{c}{20}&\multicolumn{2}{c}{5}&\multicolumn{2}{c}{10}&\multicolumn{2}{c}{20}\\\cline{3-4}\cline{5-6}\cline{7-8}\cline{9-10}\cline{11-12}\cline{13-14}
  & \multicolumn{1}{r}{$A$} & 5 & 6 & 5 & 6 & 5 & 6 & 5 &6 & 5 & 6 & 5 & 6\\
\hline
DL$_{1/n}$ & original & 42.47 & 17.51 & 82.95 & 34.45 & 143.55 & 65.15 & 108.32 & 44.32 & 181.16 & 79.22 & 316.33 & 129.44\\
DL$_{1/n}$ & corrected & 16.37 & 7.59 & 35.37 & 14.87 & 66.81 & 32.42& 38.19 & 15.40 & 73.99 & 33.43 & 140.61 & 66.51\\
DL$_{1/n}$ & alternative & 14.22 & 6.96 & 28.24 & 14.27 & 61.40 & 29.99 & 30.56 & 14.53 & 64.57 & 31.98 & 126.77 & 63.30\\
\hline
DL$_{1/2}$ & original & 12.47 & 11.19 & 20.31 & 17.73 & 36.83 & 31.62 & 25.22 & 22.69 & 41.81 & 36.60 & 74.11 & 63.31\\
DL$_{1/2}$ & corrected & 13.98 & 13.18 & 20.42 & 18.86 & 34.10 & 30.90 & 28.15 & 26.59 & 41.98 & 38.64 & 68.56 & 61.70\\
DL$_{1/2}$ & alternative & 13.98 & 13.25 & 20.44 & 18.84 & 34.17 & 30.84 & 28.18 & 26.49 & 41.89 & 38.51 & 68.67 & 61.75\\
\hline
\end{tabular}
}
\end{table}

\begin{table}[tp]
\caption{Squared error comparison over $100$ replicates. Average squared error across replicates reported for DL with $a=1/n$ and $a=1/2$ based on the original, the corrected and the alternative algorithm. Signal strength $A=7$ and $A=8$.}
\label{tab:sim2}
\centering
\resizebox*{\textwidth}{!}{
\begin{tabular}[t]{rl!{\extracolsep{6pt}}r!{\extracolsep{0pt}} r !{\extracolsep{6pt}}r!{\extracolsep{0pt}} r!{\extracolsep{6pt}}r!{\extracolsep{0pt}} r !{\extracolsep{6pt}}r!{\extracolsep{0pt}} r !{\extracolsep{6pt}}r!{\extracolsep{0pt}} r !{\extracolsep{6pt}}r!{\extracolsep{0pt}} r}
\hline\hline
& \multicolumn{1}{r}{$n$}&\multicolumn{6}{c}{100} & \multicolumn{6}{c}{200}\\
\cline{3-8}\cline{9-14}
&\multicolumn{1}{r}{$\frac{q_n}{n}$\%}&\multicolumn{2}{c}{5}&\multicolumn{2}{c}{10}&\multicolumn{2}{c}{20}&\multicolumn{2}{c}{5}&\multicolumn{2}{c}{10}&\multicolumn{2}{c}{20}\\\cline{3-4}\cline{5-6}\cline{7-8}\cline{9-10}\cline{11-12}\cline{13-14}
  & \multicolumn{1}{r}{$A$} & 7 & 8 & 7 & 8 & 7 & 8 & 7 & 8 & 7 & 8 & 7 & 8\\
\hline
DL$_{1/n}$ & original & 7.43 & 5.96 & 14.98 & 12.86 & 32.47 & 25.90& 18.51 & 12.65 & 35.69 & 26.90 & 66.90 & 52.19\\
DL$_{1/n}$ & corrected & 5.72 & 5.40 & 11.75 & 11.22 & 25.67 & 23.58& 11.90 & 11.41 & 26.25 & 24.03 & 51.09 & 46.98\\
DL$_{1/n}$ & alternative & 5.73 & 5.42 & 11.73 & 11.16 & 25.15 & 23.54& 11.93 & 11.38 & 25.88 & 23.85 & 50.02 & 46.87\\
\hline
DL$_{1/2}$ & original & 10.69 & 10.41 & 16.74 & 16.19 & 29.34 & 28.22 & 21.66 & 21.10 & 34.34 & 33.20 & 58.73 & 56.40\\
DL$_{1/2}$ & corrected & 12.83 & 12.63 & 18.25 & 17.84 & 29.41 & 28.67& 25.98 & 25.56 & 37.23 & 36.35 & 58.83 & 57.16\\
DL$_{1/2}$ & alternative & 12.90 & 12.66 & 18.25 & 17.88 & 29.39 & 28.64 & 25.97 & 25.60 & 37.17 & 36.37 & 58.76 & 57.13\\
\hline
\end{tabular}
}
\end{table}

In this section we provide results of the simulation study in Section 4
in \citet{dl} based on both correct algorithms. Table
\ref{tab:sim1} and Table \ref{tab:sim2} in this manuscript refer to
Table 1 and Table 2 in the original manuscript. Differences between the
results in the original manuscript and the results presented here using
the original algorithm are probably due to the use of different software
packages. In general, the scores with respect to the original algorithm align
with those presented in \citet{dl} except for
DL\(_{1/n}\) in the scenarios where \(A \in \{5,6\}\), i.e.~in the
scenarios with relatively weak signals. For DL\(_{1/n}\), the scores
based on the correct algorithms are better for all considered
dimensionalities, sparsity and signal levels compared to the scores
based on the original algorithm. For DL\(_{1/2}\), the scores based on
the correct algorithms are a little worse than the scores based on the
original algorithm when the model size \(q_n\) is small or when the
signal level \(A\) is high. Overall, the differences are greater for
DL\(_{1/n}\).

\begin{table}[tp]
\caption{Squared error comparison over $100$ replicates. Average squared error for the posterior median reported for DL with $a=1/n$ and $a=1/2$ based on the original, the corrected, and the alternative algorithm.}
\label{tab:sim3}
\centering
\begin{tabular}[t]{rlrrrrrr}
\hline\hline
\multicolumn{2}{c}{$n$}&\multicolumn{6}{c}{1000}\\ 
\cline{3-8}
\multicolumn{2}{c}{$A$}& 2 & 3 & 4 & 5 & 6 & 7\\
\hline
DL$_{1/n}$ & original & 384.14 & 795.03 & 1106.57 & 845.25 & 387.41 & 202.93\\
DL$_{1/n}$ & corrected & 365.64 & 667.63 & 671.93 & 381.47 & 182.34 & 132.72\\
DL$_{1/n}$ & alternative & 358.21 & 614.98 & 572.78 & 316.10 & 165.99 & 129.65\\
\hline
DL$_{1/2}$ & original & 268.00 & 314.29 & 266.72 & 215.07 & 189.34 & 178.42\\
DL$_{1/2}$ & corrected & 277.08 & 304.44 & 256.85 & 217.63 & 200.76 & 193.79\\
DL$_{1/2}$ & alternative & 277.23 & 304.57 & 257.24 & 217.54 & 200.77 & 193.86\\
\hline
\end{tabular}
\end{table}

Table \ref{tab:sim3} depicts the results for the high-dimensional
simulation setting and refers to Table 3 in the original manuscript.
Similarly, for DL\(_{1/2}\), the differences between the
results based on the original algorithm and the correct algorithms are
relatively small, whereas for DL\(_{1/n}\) the differences are
considerable.
For most considered scenarios, the scores based on the corrected
algorithm and the scores based on the alternative algorithm align almost perfectly. We observe differences in the scenarios where, for
DL\(_{1/n}\), the squared error loss is strikingly high, though.
In these cases, the squared error loss is always the smallest for the
alternative algorithm, which seems to be more robust for very small
values of the parameter \(a\).
We hypothesize that this is to do with numerical and mixing issues, and leave a more detailed investigation for further research.

\bibliographystyle{apalike}
\bibliography{references}

\end{document}